\documentclass[aps,prl,reprint,superscriptaddress,longbibliography,footinbib]{revtex4-2}
\usepackage{amsmath,amssymb,bm,graphicx,xcolor,hyperref}
\usepackage[T1]{fontenc}
\newcommand{\dd}{\mathrm d}
\DeclareMathVersion{fkdisplay}
\AtBeginDocument{\SetSymbolFont{operators}{fkdisplay}{OT1}{cmr}{m}{n}}
\newcommand{\sumint}{\mathop{\vcenter{\hbox{\ooalign{%
  \hfil$\displaystyle\sum$\hfil\cr
  \hfil$\displaystyle\int$\hfil\cr}}}}\displaylimits}
\hypersetup{colorlinks=true,linkcolor=blue,citecolor=blue,urlcolor=blue}
\allowdisplaybreaks[0]
\usepackage{booktabs}

\makeatletter
\newcommand{\equationalias}[2]{%
  \AtEndDocument{%
    \@ifundefined{r@#2}{}{%
      \begingroup
      \expandafter\let\expandafter\alias@data\csname r@#2\endcsname
      \protected@write\@auxout{}{\string\newlabel{#1}{\alias@data}}%
      \endgroup
    }%
  }%
}
\makeatother

\begin{document}
\title{Time Advances and Dispersive Constraints in Dressed Gravitational Scattering}
\hypersetup{pdftitle={Time Advances and Dispersive Constraints in Dressed Gravitational Scattering},pdfauthor={Ziyu Dong}}
\author{Ziyu Dong}
\email{zdong@ictp.it}
\affiliation{The Abdus Salam International Centre for Theoretical Physics, Strada Costiera 11, 34151 Trieste, Italy}
\begin{abstract}
We derive a low energy group time advance in 4D dressed gravitational scattering at leading eikonal order with full soft dependence for a sharp dressing profile of finite soft scale. Dispersive matching of nonnegative absorption excludes the amplitude's analytic crossing with high energy boundedness. This reflects internal scales growing with energy via string diffusion in quantum gravity and high energy nonlinear graviton rescattering, coupled to soft clouds beyond leading soft order. Interference preserving joint detection of hard particles and coherent soft radiation may reveal the advance.
\end{abstract}
\maketitle

In four dimensional gravity, exclusive Fock state amplitudes are infrared (IR) divergent because hard particles are inseparable from arbitrarily soft gravitational fields \cite{Weinberg:1965nx}. Inclusive cross sections remove this divergence by summing over unobserved radiation. Alternatively, Faddeev--Kulish (FK) dressing attaches coherent soft graviton clouds to asymptotic states \cite{Chung:1965,Kibble:1968asymptotic,Kibble:1968scattering,Kulish:1970ut,Ware:2013zja}, defining an IR finite, unitary S matrix \cite{HannesdottirSchwartz:2020} that preserves interference between incoming superpositions \cite{Carney:2018soft}. IR finiteness, however, does not guarantee the analytic crossing structure and high energy bounds used for exclusive scattering.

With a finite soft scale \(\omega\), a dressed amplitude samples an entire soft band, whereas soft theorems and asymptotic symmetries constrain the zero frequency limit \cite{Weinberg:1965nx,HeLysovMitraStrominger:2014,CachazoStrominger:2014}. These wavelengths can resolve hard trajectory separation, Fourier conjugate to transverse exchange momentum, and the spatial extent of high energy internal dynamics. The amplitude's long range response therefore depends on interactions between the soft dressing background and these internal dynamics.

We compute the low energy dressed \(2\to2\) amplitude at leading eikonal (LE) order, retaining the full soft dependence on trajectory separation. Narrow packets separated on the soft wavelength scale can exhibit an IR finite group time advance. A twice subtracted dispersion relation (2SDR) connects its negative low energy coefficient to nonnegative absorption in both physical channels and a contour at complex infinity. Dispersive bootstrap \cite{AdamsEtAl2006,Caron-Huot:2021rmr,Caron-Huot:2022ugt} therefore excludes simultaneous analytic crossing and high energy (HE) boundedness for the UV dressed amplitude, as assumed for exclusive 2SDRs \cite{Haring:2022cyf}.

To understand this obstruction, we examine how the soft cloud couples to UV dynamics. The spectrum includes graviton exchanges resummed from higher loops \cite{Ciafaloni:2019infrared,DiVecchiaHeissenbergRussoVeneziano_GravEikonal}, higher spin resonances in quantum gravity (QG) \cite{Haring:2024wyz}, and eventually black hole formation \cite{AmatiCiafaloniVeneziano_TransplanckianString,GiddingsPorto:2010}. String diffusion and nonlinear GR rescattering show how growing internal scales invalidate the external soft factor description. Nonlinear effects already matter in small deflection scattering before strong gravity sets in. Strong gravity also involves black hole radiation and horizon soft degrees of freedom \cite{HawkingPerryStrominger:2016}. Comparing soft dressing with AdS$_4$ curvature shows how different IR regulating backgrounds affect this UV response.

\paragraph*{Finite soft FK dressing.---}
For dressed \(2\to2\) scattering in four dimensions, a timelike unit vector \(n\) and sharp soft profile \(f_\omega(|\bm k|_n)=\Theta(\omega-|\bm k|_n)\), where \(|\bm k|_n=\sqrt{(n\cdot k)^2-k^2}\), specify the dressing background. The profile restricts spatial momenta of real and virtual soft gravitons emitted or absorbed by each hard particle. Each particle and cloud form an asymptotic state with finite cloud energy but divergent soft occupation \cite{Arkani-Hamed:2020gyp}. Coherent sums over soft multiplicities and phase space integrals cancel the IR divergence, while \(\omega\to0\) recovers the singular exclusive limit. The isotropic profile preserves translations and spatial \(SO(3)\), but selects a boost frame, so Lorentz covariance requires transforming \(n\) with the hard momenta.

FK dressing constructs these states through gravitational Wilson lines along recoil-free asymptotic half rays, acting on hard Fock states at \(t=0\) in the interaction picture \cite{Ware:2013zja,White:2011,MelvilleEtAl:2013}. Their stress-energy current couples to the dressing field through \(V_{\rm as}(t)=(\kappa/2)\int\dd^3\bm x\,T_{\rm as}^{\mu\nu}h_{\mu\nu}^{<}\), with \(\kappa^2=32\pi G_N\). In the rest frame of \(n\), where \(|\bm k|_n=|\bm k|\), the leading current and soft field are
\begin{equation}
 \begin{aligned}
 &T_{\rm as}^{(0),\mu\nu}(t,\bm x)=\int\dd\Pi_p\,\rho(p)\frac{p^\mu p^\nu}{p^0}
 \delta^{(3)}\Bigl(\bm x-\frac{\bm p}{p^0}t\Bigr),\\
 &h_{\mu\nu}^{<}(x)=\int\dd\Pi_k\,f_\omega(|\bm k|)\,
 \bigl[\epsilon_{\mu\nu}^{h}(k)a_h(k)e^{-ik\cdot x}+\mathrm{h.c.}\bigr].
 \end{aligned}
 \label{PRL:as-potential}
\end{equation}
\noindent Here \(\dd\Pi_p=\dd^3\bm p/[(2\pi)^3 2p^0]\) is the positive energy on-shell measure, likewise for \(k\), and \(\rho(p)=\sum_\sigma b_\sigma^\dagger(p)b_\sigma(p)\) sums over discrete hard states. The half ray through the origin yields the Fourier current \(T_i^{(0),\mu\nu}(k)=p_i^\mu p_i^\nu/(p_i\cdot k-i0)\), generating the external line leading soft factor \cite{Weinberg:1965nx}. Dressing implements inverse asymptotic evolution from \(t=0\) to the past and from the future to \(t=0\) \cite{HannesdottirSchwartz:2020},
\begingroup
\thickmuskip=2mu
\renewcommand{\mintagsep}{2pt}
\begin{gather}
 U_{\rm as}(-\infty,0)=\overline{\mathcal T}e^{i\int_{-\infty}^{0}V_{\rm as}\dd t},\enspace
 U_{\rm as}(0,\infty)=\overline{\mathcal T}e^{i\int_{0}^{\infty}V_{\rm as}\dd t}.
 \label{PRL:inverse}
\end{gather}
\endgroup
\noindent Although \(h^{<}\) in \(V_{\rm as}\) is free, inverse time ordering generates off-shell propagators through its contractions. Combining the asymptotic evolutions with the Dyson operator \(\mathcal S_D\) gives the dressed S matrix and its optical theorem for a hard Fock state \(|i\rangle\),
\begin{equation}
 \begin{aligned}
 &\mathcal S_{\rm dr}=U_{\rm as}(0,\infty)\,\mathcal S_D\,U_{\rm as}(-\infty,0)=1+i\mathcal T_{\rm dr},\\
 &2\operatorname{Im}\langle i|\mathcal T_{\rm dr}|i\rangle
 =\mathop{\smash[b]{\sumint}}\nolimits_X|\langle i|\mathcal T_{\rm dr}|X\rangle|^2\ge0.
 \end{aligned}
 \label{PRL:unitarity}
\end{equation}
\noindent The second line follows by inserting complete intermediate states into \(\mathcal S_{\rm dr}\mathcal S_{\rm dr}^{\dagger}=1\), since \(\mathcal S_D\) and \(U_{\rm as}\) are unitary, the latter by Hermiticity of \(V_{\rm as}\).

For the leading current, the Magnus expansion factorizes \(U_{\rm as}\) into the coherent radiation factor \(e^R\) and virtual Coulomb phase \(e^{i\Phi}\) \cite{Ware:2013zja}, as detailed in Supplemental Material (SM) \cite{SupplementalMaterial}. Since \(R\) retains soft creation and annihilation operators, its cloud need not commute through long range interactions in \(\mathcal S_D\). The interaction picture Hamiltonian \(H_{\rm int}\) makes this coupling explicit in the full evolution,
\begingroup
\mathversion{fkdisplay}
\thickmuskip=2mu
\medmuskip=2mu
\renewcommand{\mintagsep}{2pt}
\begin{gather}
 \mathcal S_{\rm dr}=\mathcal T\exp\Bigl[-i\int U_{\rm as}(0,t)(H_{\rm int}-V_{\rm as})(t)U_{\rm as}(t,0)\,\dd t\Bigr].
 \label{PRL:asymptotic-evolution}
\end{gather}
\endgroup
\noindent The remainder \(H_{\rm int}-V_{\rm as}\) contains UV interactions and nonlinear graviton self-interactions beyond the linear asymptotic coupling, and generally does not commute with \(V_{\rm as}\).

\equationalias{PRL:remainder}{PRL:asymptotic-evolution}
At leading soft (LS) order, weak commutation moves clouds between incoming and outgoing sides, retaining the induced soft contractions \cite{Choi:2017ylo}. These yield a virtual soft factor fixed by the hard data and dressing profile. For four point scattering, its exponent is defined by the external currents and a real function \(g\),
\begingroup
\thinmuskip=2mu
\medmuskip=2mu
\thickmuskip=2mu
\renewcommand{\mintagsep}{2pt}
\begin{gather}
 K[g]=-\frac{i\kappa^2}{8}\sum_{i,j=1}^{4}\int
 \frac{\dd^4k\,g(k)T_i^{(0),\mu\nu}P_{\mu\nu\rho\sigma}T_j^{(0),\rho\sigma *}}
 {(2\pi)^4(k^2+i0)}.
 \label{PRL:virtual-kernel}
\end{gather}
\endgroup
\noindent with all hard momenta incoming and \(P_{\mu\nu\rho\sigma}\) the de Donder propagator numerator. Cloud--cloud contractions carry \(f_\omega^2\), conjugated by inverse time ordering. Emission by a cloud and absorption by the hard process, or vice versa, instead carries \(f_\omega\). Acting together on the exclusive matrix element \(S_{\rm ex}\), they give
\begin{equation}
 S_{\rm dr}\overset{\rm LS}{\simeq}e^{K^*[f_\omega^2]-K[f_\omega]-K^*[f_\omega]}S_{\rm ex}=e^{-K[f_\omega]}S_{\rm ex}.
 \label{PRL:virtual-cancellation}
\end{equation}
\noindent where the last equality uses \(f_\omega^2=f_\omega\). The cloud--hard contractions assume recoil-free emission or absorption throughout the soft band, requiring external hardness \(|s|,|t|,|u|\gg\omega^2\). Here \(s\) is the squared hard center-of-mass energy. For massless four point scattering, the exponent becomes \cite{Weinberg:1965nx,NaculichSchnitzer:2011,White:2011}
\begingroup
\thinmuskip=2mu
\medmuskip=2mu
\thickmuskip=2mu
\renewcommand{\mintagsep}{2pt}
\begin{gather}
 K[f_\omega]=-\frac{2G_N}{\pi}\ln\frac{\omega}{\lambda_{\rm IR}}
 [s\ln(-s)+t\ln(-t)+u\ln(-u)],
 \label{PRL:virtual-factor}
\end{gather}
\endgroup
\noindent whose apparent frame independence follows from the LS integrand's degree \(-4\) homogeneity in \(k\) in eq.~\eqref{PRL:virtual-kernel}. Defining both cutoffs in the same frame \(n\) leaves only \(\ln(\omega/\lambda_{\rm IR})\) after radial integration. Trajectory positions or extended UV dynamics can spoil this property beyond LS.

\paragraph*{Beyond Leading Soft Order.---}
At LE order, \(sb^2\gg1\), the impact parameter representation expresses the elastic matrix element through its eikonal phase as \(S=e^{2i\delta}\). At low energy and large separation, massive exchange is suppressed, leaving conservative first post-Minkowskian (1PM) scattering \cite{DiVecchiaHeissenbergRussoVeneziano_GravEikonal}. Equation~\eqref{PRL:virtual-factor} then reduces to \(\pm2iG_Ns\ln(\omega/\lambda_{\rm IR})\) for \(s\pm i0\), yielding \(\delta_{\rm dr,LS}=\delta_{\rm 1PM}-K[f_\omega]/(2i)\). Thus the soft scale acts as an effective IR cutoff in the low energy LE and LS limits.

Now take neutral narrow packets with \(b\gtrsim\omega^{-1}\) and \(\omega\) below all nonzero Standard Model masses. Extrapolating the LS phase to these separations gives a negative result that decreases without bound as \(b\) grows, violating the positive time delay criterion inferred from causality for exclusive scattering \cite{Camanho:2014apa}. Retaining the current's full spatial dependence therefore requires going beyond LS order. The first correction is \cite{CachazoStrominger:2014,Sen:2017soft,Choi:2019rlz}
\begingroup
\thinmuskip=2mu
\medmuskip=2mu
\thickmuskip=2mu
\renewcommand{\mintagsep}{2pt}
\begin{gather}
 T_i^{(1),\mu\nu}=-i\frac{p_i^{(\mu}k_\rho J_i^{\rho\nu)}}{p_i{\cdot}k-i0}
 =-i(k{\cdot}b_i)T_i^{(0),\mu\nu}+ip_i^{(\mu}b_i^{\nu)},
 \label{PRL:subleading-current}
\end{gather}
\endgroup
\noindent where \(J_i^{\mu\nu}=2b_i^{[\mu}p_i^{\nu]}\) is the orbital angular momentum of a straight trajectory displaced by \(b_i^\mu\). Contracting with the other line's LS current through \(P_{\mu\nu\rho\sigma}\) leaves only the linear shift of \(T_i^{(0)}\), giving relative displacement \(k\cdot(b_1-b_2)\). Resumming these shifts at finite \(\omega b\) retains the currents' full transverse position dependence, while LE scattering permits neglecting recoil along nearly straight trajectories. In low energy GR, these currents couple linearly to a free field, and the sharp profile separates disjoint momentum bands. Hence the soft potential commutes with the remaining interaction at all times,
\begingroup
\thinmuskip=2mu
\medmuskip=2mu
\thickmuskip=2mu
\renewcommand{\mintagsep}{2pt}
\begin{gather}
 [V_{\rm as}(t),(H_{\rm int,1PM}-V_{\rm as})(t')]=0,\notag\\
 \mathcal S_{\rm dr,1PM}\simeq
 \mathcal T\exp\left[-i\int\dd t\,(H_{\rm int,1PM}-V_{\rm as})(t)\right].
 \label{PRL:commute}
\end{gather}
\endgroup
\noindent Here \(H_{\rm int,1PM}\) couples the two eikonal trajectories through the unfiltered graviton field. The subtraction removes \(q<\omega\) from each 1PM ladder rung's transverse momentum integral, where \(q^2=-t\). For \(n\) in the longitudinal plane, resummation gives the dressed LE phase
\begin{equation}
 \frac{\delta_{\rm dr}(s,b)}{G_Ns}
 =\int_\omega^\infty\frac{\dd q}{q}J_0(qb)\equiv G_\omega(b).
 \label{PRL:phase}
\end{equation}
\noindent Unlike stripping the external virtual soft factor, this subtraction retains \(\exp(i\vec q\cdot\vec b)\) and full soft spatial dependence. It applies only to low energy LE scattering, neglecting nonlinear gravitational and QG corrections.

\begin{figure}[t]
 \centering
 \includegraphics[width=\columnwidth]{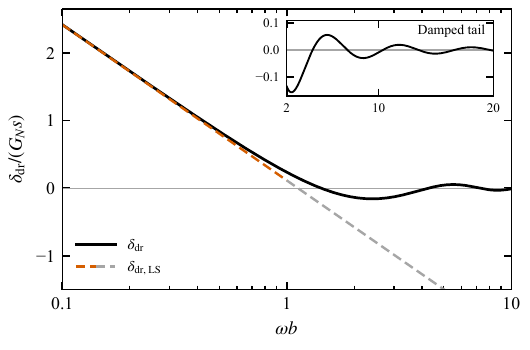}
 \caption{Spatially resolved dressing changes the long distance phase. Black shows \(\delta_{\rm dr}\); orange and gray show the controlled short distance segment at \(\omega b\ll1\) and the uncontrolled extrapolation of the same LS order phase \(\delta_{\rm dr,LS}\). The inset resolves the decaying oscillations and negative lobes.}
 \label{PRL:fig:gr}
\end{figure}

Instead of the unbounded LS logarithm, the spatially resolved phase oscillates towards zero (Fig.~\ref{PRL:fig:gr}). The sharp soft band makes the dressing background a spatial high-pass medium, analogous to Fourier-optical filtering \cite{Tanriover:2023highpass}. A narrow packet centered on a negative lobe can therefore resolve a time advance.

\paragraph*{Time advance and dispersive constraint.---}
To extract this time advance, separate the longitudinal and transverse wave functions at LE order. Identical wave functions for corresponding in- and out-states cancel the longitudinal plane wave phases, leaving a normalized relative transverse profile \(\psi(\vec b)\). For counterpropagating massless particles with \(s=4E_1E_2\), its elastic matrix element and group time delay in the low energy weak gravity regime \(G_Ns\ll1\) are \cite{Wigner:1955delay,Smith:1960lifetime}
\begingroup
\thinmuskip=2mu
\medmuskip=2mu
\thickmuskip=2mu
\renewcommand{\mintagsep}{2pt}
\begin{gather}
 \begin{aligned}
 &S_{\rm dr,\psi}(s)=\int\dd^2\vec b\,|\psi(\vec b)|^2e^{2i\delta_{\rm dr}(s,b)},\\
 &\Delta t_\psi=\operatorname{Re}\left[-i\frac{\partial_{E_1}S_{\rm dr,\psi}}{S_{\rm dr,\psi}}\right]
 \simeq8G_NE_2\int\dd^2\vec b\,|\psi|^2G_\omega(b).
 \end{aligned}
 \label{PRL:packet}
\end{gather}
\endgroup
\noindent For a Gaussian \mbox{\(|\psi|^2=e^{-|\vec b-\vec b_0|^2/(2\sigma_b^2)}/(2\pi\sigma_b^2)\)} with \(s^{-1/2}\ll\sigma_b\ll\omega^{-1}\), choosing \(\omega b_0=3\) and \(\omega\sigma_b=0.1\) gives \(\Delta t_\psi\simeq-1.01G_NE_2\). This gives an IR finite advance in the specified dressing background. The packet peak is advanced, not the wavefront determining the earliest causal signal. If a detector resolves dressing modes below \(\omega\), the measured process involves hard particles scattering in that resolved soft background, rather than hard particles alone.

Dispersive matching also requires a positive UV absorption functional for this packet. At LE order, partial waves \(d^J_{hh}(\theta)\) in a fixed helicity conserving channel reduce to \(J_0(qb_J)\), with \(b_J=(2J+1)/\sqrt{s}\) \cite{Hoffmann:2017wigner}. The packet's transverse weight is
\begin{equation}
 \begin{aligned}
 &\int_0^\infty\frac{q\,\dd q}{2\pi}
 e^{-\sigma_b^2q^2/2}J_0(qb_0)J_0(qb_J)\\
 &\qquad=\frac{e^{-(b_J^2+b_0^2)/(2\sigma_b^2)}I_0(b_Jb_0/\sigma_b^2)}{2\pi\sigma_b^2}\ge0,
 \end{aligned}
 \label{PRL:positive}
\end{equation}
\noindent where \(I_0\) is the modified Bessel function of the first kind. With this nonnegative weight, the optical theorem in eq.~\eqref{PRL:unitarity} ensures nonnegative physical absorption in both the \(s\) and \(u\) channels at LE order.

Packet separation suppresses short range and contact terms, while neutrality excludes Coulomb exchange. Matching uses a low energy contour within the LE window \(|s|\sigma_b^2\gg1\), \(G_N|s|\ll1\), rather than taking \(s\to0\) at fixed packet width. A compatible UV completion must reproduce the negative coefficient in eq.~\eqref{PRL:packet}\footnote{For \(\omega\ll\Lambda_{\rm EFT}\), cubic operators such as \(RFF\) and \(R^3\) give \(\mathcal O(\omega^2/\Lambda_{\rm EFT}^2)\) power corrections to this packet \cite{Camanho:2014apa}; short range exchange and local four point terms are exponentially suppressed.}, which Cauchy contour deformation relates to all UV cuts and poles and the outer contour at infinity,
\begingroup
\renewcommand{\mintagsep}{2pt}
\begin{gather}
 \frac{\Delta t_\psi}{2E_2}
 \simeq\oint_{\rm low}\frac{\dd s\,\mathcal M_{\rm dr,\psi}}{2\pi i s^3}
 =\Bigl[\oint_{\rm cut}+\oint_\infty\Bigr]\frac{\dd s\,\mathcal M_{\rm dr,\psi}}{2\pi i s^3},
 \label{PRL:dispersion}
\end{gather}
\endgroup
\noindent where \(\mathcal M_{\rm dr,\psi}=-2is(S_{\rm dr,\psi}-1)\) on the physical \(s\) channel boundary. To identify the cuts with physical absorption, we require \emph{analytic crossing} at all UV scales: (i) real analyticity, \(\mathcal M_{\rm dr,\psi}(s^*)=\mathcal M_{\rm dr,\psi}(s)^*\), identifying the nonnegative physical \(\operatorname{Im}\mathcal M_{\rm dr,\psi}\) with \(\operatorname{Disc}\mathcal M_{\rm dr,\psi}/(2i)\); (ii) crossing symmetry, identifying the \(s\) and \(u\) channels as boundary values of one analytic function. If HE boundedness, \(\mathcal M_{\rm dr,\psi}/s^2\to0\) uniformly at complex infinity, also held, the outer contour would vanish and contradict the negative low energy side. Thus the two properties cannot coexist in a matching dressed amplitude. By contrast, for exclusive packets, the real axis unitarity bound extends to complex directions by the Phragm\'en--Lindel\"of principle under analyticity and subexponentiality \cite{Caron-Huot:2022ugt,Dong:2026halfspace}.

This obstruction concerns the UV completion, not the low energy GR kernel. At LE order, the latter is \(\mathrm S_{\rm dr,1PM}=\exp\{-2[\log(-s)-\log s]\delta_{\rm dr}/\pi\}\) and satisfies analytic crossing, without establishing this property for the full packet amplitude \(\mathcal M_{\rm dr,\psi}\) at UV scales. At higher energies, soft cloud interactions with internal long range dynamics invalidate the commutation approximation in eq.~\eqref{PRL:commute}, so extending the dressed amplitude requires more than linear GR coupling to external trajectories.

\begin{figure*}[t]
 \centering
 \includegraphics[width=\textwidth]{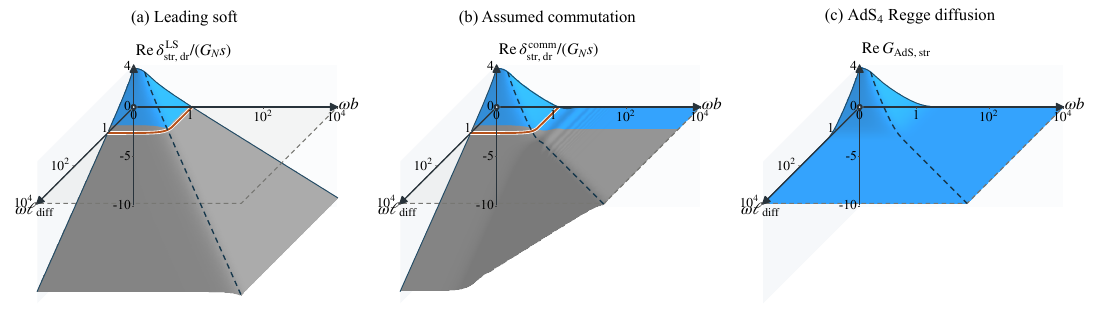}
 \caption{Eikonal phases versus impact parameter \(b\) and effective string size \(\ell_{\rm diff}\). Gray regions mark uncontrolled extrapolations of LS stripping in (a) and assumed commutation in (b), as defined in eq.~\eqref{PRL:bare-string}. Panel (c) gives the AdS\(_4\) comparison in eq.~\eqref{PRL:ads}. The flat panels use \(\alpha'\omega^2=10^{-4}\).}
 \label{PRL:fig:diffusion}
\end{figure*}

\paragraph*{Internal scales at transplanckian energies.---}
The external hardness condition \(|s|,|t|,|u|\gg\omega^2\) is only pseudo-hardness: it does not exclude UV effects on long range interactions. QG excitations and nonlinear GR self-interactions can generate spatial scales that grow with energy even while the external particles remain hard. We first illustrate this with string diffusion.

For \(-t=q^2\ll s\), the type-II dilaton amplitude has the leading Regge form \cite{Haring:2024wyz}
\begin{equation}
 \mathcal M_{\rm VS}(s,-q^2)
 =\frac{8\pi G_Ns^2}{q^2}e^{-Y_cq^2/4}
 \frac{\Gamma(1+\alpha'q^2/4)}{\Gamma(1-\alpha'q^2/4)}.
 \label{PRL:string}
\end{equation}
\noindent Here \(Y_c=2\alpha'\log(-i\alpha's/4)=\ell_{\rm diff}^2-i\pi\alpha'\). The diffusion length \(\ell_{\rm diff}\) describes the collective, energy averaged contribution of \(s\) channel string resonances, dual to \(t\) channel Reggeized graviton exchange \cite{DiVecchiaHeissenbergRussoVeneziano_GravEikonal,BrowerPolchinskiStrasslerTan_BPST_Pomeron}. For a graviton of frequency \(\nu\) with the SM emission direction and polarization, the leading and subleading soft factors \(S^{(0)}\simeq-i\kappa q/\nu\) and \(S^{(1)}\simeq\kappa(q\partial_q-2s\partial_s)/4\), acting in the Regge limit \(q^2\ll\alpha'^{-1}\ll s\), give
\begin{equation}
 \frac{S^{(1)}\mathcal M_{\rm VS}}{S^{(0)}\mathcal M_{\rm VS}}
 =-\frac{i\nu}{q}\left[\frac32+\frac{q^2Y_c}{8}+O(\alpha'q^2)\right].
 \label{PRL:stringratio}
\end{equation}
\noindent Even with \(\nu\le\omega\ll q\), the correction is enhanced by \(\nu q\ell_{\rm diff}^2\). Three spatial scales are therefore relevant. At \(b\gtrsim\omega^{-1}\), the cloud resolves the external separation, invalidating LS even in low energy GR. When \(\ell_{\rm diff}\) reaches \(b\), the other particle resolves collective string exchange. Once \(\ell_{\rm diff}\gtrsim\omega^{-1}\), the cloud itself interacts with the extended string structure, i.e., higher soft orders are no longer suppressed and the commutation approximation in eq.~\eqref{PRL:commute} fails for either \(b\gtrless\omega^{-1}\).

To compare prescriptions, temporarily neglect nonlinear GR corrections and eikonalize the two-particle-irreducible tree kernel of the LE ladders. In the Regge limit, the exclusive phase and dressed phases from LS stripping and assumed commutation are
\begingroup
\thinmuskip=1.25mu
\medmuskip=0mu
\thickmuskip=0.5mu
\renewcommand{\mintagsep}{0.5pt}
\begin{gather}
 \{\delta_{\rm str},\delta_{\rm str,dr}^{\rm LS},\delta_{\rm str,dr}^{\rm comm}\}
 =\{\delta_{\rm 1PM},\delta_{\rm dr,LS},\delta_{\rm dr}\}
 -\frac{\mkern-2mu G_Ns\mkern-2mu}{2}E_1\bigl(\frac{\mkern-2mu b^2\mkern-2mu}{\mkern-2mu Y_c\mkern-2mu}\bigr),
 \label{PRL:bare-string}
\end{gather}
\endgroup
\equationalias{PRL:subtractions}{PRL:bare-string}
\par\noindent where \(E_1\) is the exponential integral. The LS prescription strips the virtual soft factor in eq.~\eqref{PRL:virtual-factor}. Its formal extrapolation violates the subexponentiality used to derive HE boundedness. Assumed commutation instead removes only the soft region of \(t\) channel graviton exchange, retaining the full low energy spatial dependence and time advance. Neither prescription includes the coupled UV--cloud evolution in eq.~\eqref{PRL:asymptotic-evolution}, limiting their validity as shown in Fig.~\ref{PRL:fig:diffusion}.

Note that the low transfer part of the full kernel in eq.~\eqref{PRL:string} cannot be cut off, because the kernel describes collective string exchange, not a single \(t\) channel graviton. Such a cutoff distorts the string resonance contribution and produces regions with \(\operatorname{Im}\delta<0\), violating unitarity after exponentiation.

Changing the background also illustrates consistent IR regulation at low and high energies. AdS\(_4\) curvature regulates the IR through bulk packet scattering or boundary correlators instead of a flat spacetime S matrix. For eikonal scattering of two single particle wave packets \cite{Cornalba:2006xk,Cornalba:2007eikonal,Brower:2007eikonal}, transverse graviton exchange is governed by the curvature-shifted Laplace--Beltrami operator \(\mathcal L_{\rm AdS}=-\Delta_{H_2}+2/R^2\) on hyperbolic space \(H_2\) of radius \(R\). With the logarithmic normalization of \(G_\omega\),
\begin{equation}
 \begin{aligned}
 &G_{\rm AdS}(b)=\cosh(b/R)\log\coth\frac{b}{2R}-1,\\
 &G_{\rm AdS,str}(s,b)=e^{-Y_c\mathcal L_{\rm AdS}/4}G_{\rm AdS}(b).
 \end{aligned}
 \label{PRL:ads}
\end{equation}
\noindent Here \(b\) is geodesic separation. The first line satisfies \(\mathcal L_{\rm AdS}G_{\rm AdS}=2\pi\delta_{H_2}\) and decays as \(e^{-2b/R}\) at large \(b\). The second approximates leading Regge diffusion with the graviton pole residue \cite{Cornalba:2007regge}, requiring weak string curvature \(\alpha'/R^2\ll1\) and small transverse transfer \(\alpha'\mathcal L_{\rm AdS}\ll1\) for contributing modes. Then \(Y_c\to\ell_{\rm diff}^2\) estimates its real part.

Matching \(\omega R=e^{1-\gamma_E}\) gives the two backgrounds the same effective IR cutoff for \(b,\ell_{\rm diff}\ll\omega^{-1},R\), but their long distance responses differ. At large \(b\), both phases vanish, but dressing has negative phase lobes giving time advances and sign changes of \(\partial_b\delta_{\rm dr}\) giving alternating attractive and repulsive deflections. AdS instead maintains positive delay and weakening attraction. The curved AdS background modifies the string diffusion heat kernel \cite{GrigoryanNoguchi:1998}, so \(\operatorname{Re}G_{\rm AdS,str}\) stays positive and tends to zero as \(\ell_{\rm diff}\) grows in the real diffusion approximation. For dressing, the full cloud--string interaction is still needed to determine whether \(\operatorname{Re}\delta\) is positive in the UV.

\begin{figure}[t]
 \centering
 \includegraphics[width=2.65in]{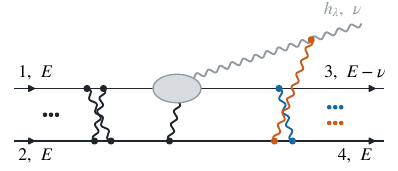}
 \caption{Ladder exchanges around a single graviton emission. Black shows the exchanges before emission and the exchange attached to the tree emission blob; blue and orange show subsequent hard particle exchange and emitted graviton rescattering. Dots indicate further exchanges, including all interleavings of blue and orange.}
 \label{PRL:fig:rescattering}
\end{figure}

Growing long range effects do not necessarily require string dynamics, since nonlinear GR self-interactions also cause graviton rescattering as \(s\) increases. For a graviton \(h_\lambda\) of helicity \(\lambda\) emitted from the 1PM ladders, factor out the elastic phase as \(\mathcal M^{\rm rad}(2\to2+h_\lambda)=e^{2i\delta_{\rm 1PM}}\mathfrak M_\lambda\) \cite{Ciafaloni:2015unified,Ciafaloni:2019infrared}. Keeping small hard deflection and forward emission, \(\nu\ll E=\sqrt{s}/2\), the leading longitudinal and transverse momenta relative to incoming \(p_1\) are \(\nu p_1/E\) and \(\nu\vec\theta\), with \(\theta\ll1\). In Fig.~\ref{PRL:fig:rescattering}, \(\vec b\) separates the outgoing hard particles and \(\vec x\) locates the graviton relative to its parent. Transverse energy-centroid balance then gives incoming separation \(\vec b-(\nu/E)\vec x\), so the phases before and after emission at fixed \(s\) are
\begin{equation}
 \begin{aligned}
 &\delta_{\rm in}=\delta_{\rm 1PM}(|\vec b-(\nu/E)\vec x|),\\
 &\delta_R=(1-\nu/E)\delta_{\rm 1PM}(b)
 +(\nu/E)\delta_{\rm 1PM}(|\vec b-\vec x|).
 \end{aligned}
 \label{PRL:rescattering}
\end{equation}
\noindent The tree emission blob includes radiation from nearly on-shell hard lines 1 and 3 and from the off-shell exchanged graviton through nonlinear self-coupling. Summing its insertion over all ladder and crossed ladder orderings combines the phases into \(e^{2i\delta_R}-e^{2i\delta_{\rm in}}\); the remaining \(\vec x\) integral gives the emission amplitude (SM). With \(R_s=2G_N\sqrt{s}\) and \(2R_s\ll b\ll\nu^{-1}\), its logarithmically enhanced form is \cite{Ciafaloni:2019infrared}
\begin{equation}
 \mathfrak M_\lambda\simeq\frac{\sqrt{G_Ns}}{4\pi i\nu}
 \left[e^{i\lambda\psi}-e^{-2i\nu R_s\log\frac{2}{b\nu\theta}}\right].
 \label{PRL:radiation-two-phases}
\end{equation}
\noindent Here \(\psi=\arg\vec\theta-\arg(\vec\theta-\vec\Theta)\), with hard deflection \(\vec\Theta=-2R_s\vec b/b^2\). Expanding for \(\nu R_s\log\frac{2}{b\nu\theta}\ll1\) recovers the leading soft factor. When this combination reaches order one, nonlinear rescattering effects must be retained.

The same issue arises at generic emission angles outside the forward region. For the SM transverse emission configuration and polarization, the classical soft graviton theorem \cite{Sahoo:2018lxl} gives the logarithmically enhanced subleading term
\begin{equation}
 S^{(1)}_{\rm gr}\simeq-\frac{i\kappa^3s}{32\pi}e^{-i\Theta}(\cos\Theta+i\sin\Theta)\log\frac{\nu_0}{\nu}.
 \label{PRL:radiationlog}
\end{equation}
\noindent where the hard reference scale \(\nu_0\) affects only finite subleading terms. The \(\cos\) term comes from hard scattering loop momenta above \(\nu\), while the \(\sin\) term comes from rescattering of the emitted graviton off hard particles through softer graviton exchange. At LE order, \(S^{(1)}_{\rm gr}/S^{(0)}_{\rm gr}\simeq(\nu R_s\sqrt{s}/2q)\log(\nu_0/\nu)\), displaying the enhancement. This subleading calculation omits the resummation in eq.~\eqref{PRL:radiation-two-phases}.

Thus external hardness alone cannot suppress nonlinear soft emission. When \(\nu R_s\sim\mathcal O(1)\), the low frequency expansion fails, and both internal exchanges and radiation require higher PM dynamics, with possible black hole formation included in the coupled soft and UV evolution.

\paragraph*{Conclusions.---}
The exclusive 2SDR links the low energy graviton pole to the UV spectrum, with IR divergences on both sides \cite{Haring:2024wyz}. Dressing, AdS\(_4\) curvature and suitable boundary conditions in locally flat spacetime \cite{Dong:2026halfspace} suppress the zero and near-zero modes responsible for long range divergences. Consistent IR regulation must also modify UV spectral contributions. String exchange, black hole solutions and their dissipative response depend on geometry \cite{BrowerPolchinskiStrasslerTan_BPST_Pomeron,HorowitzHubeny:2000}. Dressed states provide a flat spacetime background analogous to a spatial high-pass optical medium. When \(\ell_{\rm diff}\) or \(R_s\) approaches its characteristic length \(\omega^{-1}\), interactions with this background modify the UV response beyond dividing the exclusive amplitude by the linear GR virtual soft factor, which depends only on external momenta.

Whether gravitational dressing with a finite soft profile preserves analytic crossing at transplanckian energies remains open. Perturbative crossing and broader IR finite dressings beyond FK in QED \cite{Kapec:2017tkm,HiraiSugishita:2023}, gravitational cloud-moving identities \cite{Choi:2017ylo}, and studies of IR finite S-matrix analyticity \cite{HannesdottirSchwartz:2020} motivate testing this possibility. If analytic crossing holds, positive absorption for the packet in eq.~\eqref{PRL:packet} implies, at the same LE matching accuracy,
\begin{equation}
 \oint_\infty\frac{\dd s}{(2\pi i)s^3}\,\mathcal M_{\rm dr,\psi}<-0.503G_N.
 \label{PRL:negative-outer-contour}
\end{equation}

We expect hard particle measurements alone, inclusive over unresolved soft radiation, cannot isolate the advance because asymptotic evolution also specifies a coherent cloud and an FK phase \cite{Ware:2013zja,Carney:2018soft}. We propose jointly measuring energy correlators of resolvable radiation in the dressing band at future null infinity \cite{HerrmannKologluMoult:2026energy} and selected outgoing hard particle modes to characterize the coherent cloud for a given hard history and dressing profile. Coherently preparing incoming states at different impact parameters and studying their interference in these correlators could reveal the hard particles' relative group advance. At higher energies, these observables could probe the soft background's coupling to spatially extended quantum gravity dynamics and strong gravity.

\begin{acknowledgments}
\paragraph*{Acknowledgments---}
Z.\,D.\ is supported by the European Research Council under Grant Agreement No.~101039756.
\end{acknowledgments}

\bibliography{refs}

\clearpage
\onecolumngrid
\begingroup
\fontsize{10.95}{13.6}\selectfont
\setlength{\parindent}{1.2em}
\setlength{\parskip}{0pt}
\setlength{\abovedisplayskip}{11pt plus 3pt minus 6pt}
\setlength{\belowdisplayskip}{11pt plus 3pt minus 6pt}
\setlength{\abovedisplayshortskip}{0pt plus 3pt}
\setlength{\belowdisplayshortskip}{6.5pt plus 3.5pt minus 3pt}
\setcounter{equation}{0}
\setcounter{figure}{0}
\setcounter{table}{0}
\setcounter{section}{0}
\setcounter{secnumdepth}{2}
\renewcommand{\theequation}{S\arabic{equation}}
\renewcommand{\thefigure}{S\arabic{figure}}
\renewcommand{\thetable}{S\arabic{table}}
\renewcommand{\thesection}{\Roman{section}}
\renewcommand{\thesubsection}{\thesection.\Alph{subsection}}
\renewcommand{\theHequation}{supplement.\arabic{equation}}
\renewcommand{\theHfigure}{supplement.\arabic{figure}}
\renewcommand{\theHtable}{supplement.\arabic{table}}
\renewcommand{\theHsection}{supplement.\arabic{section}}
\newcommand{\lettereq}[1]{\eqref{#1}}
\makeatletter
\renewcommand{\section}{\@startsection{section}{1}{\z@}%
  {-3.5ex plus -1ex minus -.2ex}{2.3ex plus .2ex}{\normalfont\Large\bfseries}}
\renewcommand{\subsection}{\@startsection{subsection}{2}{\z@}%
  {-3.25ex plus -1ex minus -.2ex}{1.5ex plus .2ex}{\normalfont\large\bfseries}}
\let\thebibliography\NAT@thebibliography
\let\endthebibliography\endNAT@thebibliography
\let\@bibsetup\NATx@bibsetup
\renewcommand{\bibsection}{\section*{References}}
\renewcommand{\bibnumfmt}[1]{[#1]}
\renewcommand{\bibfont}{\raggedright}
\setlength{\bibsep}{0pt}
\makeatother
\pdfbookmark[0]{Supplemental Material}{supplement}
\begin{center}
{\Large\bfseries Supplemental Material}\par\medskip
{\large Time Advances and Dispersive Constraints in Dressed Gravitational Scattering}\par\medskip
Ziyu Dong
\end{center}
This supplement gives the FK radiation and phase factors, the fixed soft emission configuration, and gravitational radiation and rescattering. Equations are numbered independently of the Letter.
\section{FK dressing factors}
\label{SM:preparation}

The real dressing profile restricts spatial momentum; the virtual energy integral retains its Feynman prescription. A smooth profile must approach one at zero frequency and decay sufficiently fast at high frequency: the soft occupation and energy scale as \(\int\dd\nu\,f_\omega^2(\nu)/\nu\) and \(\int\dd\nu\,f_\omega^2(\nu)\), respectively \cite{sm-Arkani-Hamed:2020gyp}.

For the recoil-free leading current, the hard number densities commute and the free soft field commutator is a c-number. Higher nested commutators therefore vanish, so the Magnus expansion for the incoming inverse evolution terminates at second order \cite{sm-Ware:2013zja},
\begin{equation}
 \begin{aligned}
 \log U_{\rm as}(-\infty,0)&=R_{\rm in}+i\Phi_{\rm in},\\
 R_{\rm in}&=i\int_{-\infty}^{0}\dd t\,V_{\rm as}(t),\\
 i\Phi_{\rm in}&=\frac12\int_{-\infty}^{0}\dd t\int_{-\infty}^{t}\dd t'\,
 [V_{\rm as}(t),V_{\rm as}(t')].
 \end{aligned}
 \label{SM:Magnus}
\end{equation}
\noindent On a fixed hard sector, the number densities are replaced by their eigenvalues. The regulated incoming half rays give the exponents of the radiation and phase factors
\begin{equation}
 \begin{aligned}
 R_{\rm in}
 &=\frac\kappa2\sum_{i\in{\rm in},h}\int\dd\Pi_k\,f_\omega(|\bm k|)
 \left[T_i^{(0),\mu\nu}\epsilon_{\mu\nu}^{h*}a_h^{\dagger}-\mathrm{h.c.}\right],\\
 \Phi_{\rm in}
 &=\frac{\kappa^2}{16}\sum_{i,j\in{\rm in}}\int
 \frac{\dd^4k\,f_\omega^2(|\bm k|)T_i^{(0),\mu\nu}P_{\mu\nu\rho\sigma}T_j^{(0),\rho\sigma *}}
 {(2\pi)^4(k^2-i0)}+\mathrm{h.c.}
 \end{aligned}
 \label{L:displacement}
\end{equation}
\noindent The currents, polarization modes and on-shell measure follow the conventions of the Letter; the outgoing expressions are obtained analogously. \(\Phi_{\rm in}\) is a c-number, while \(R_{\rm in}\) retains the soft creation and annihilation operators. Further normal ordering the clouds of the two incoming hard particles gives
\begin{equation}
 e^{R_{\rm in}}
 =\exp\!\left[-\frac{\kappa^2}{8}\sum_h\int\dd\Pi_k\,
 f_\omega^2|\epsilon_{\mu\nu}^{h*}(T_1^{(0),\mu\nu}+T_2^{(0),\mu\nu})|^2\right]
 \mathopen{:}e^{R_{\rm in}}\mathclose{:},
 \label{SM:Rnormal}
\end{equation}
\noindent where the mixed term is the real interference between the two clouds. It comes from the soft graviton pole \(k^2=0\). The Coulomb phase instead comes from the nearly on-shell hard propagator poles, with the exchanged graviton off shell, and occurs only for two incoming or two outgoing hard lines \cite{sm-Ware:2013zja,sm-HannesdottirSchwartz:2020}.

The normal ordered operators generate coherent superpositions with arbitrary soft multiplicity on the two sides of \(\mathcal S_D\), accompanied by the real interference factors and Coulomb phases. At LS order, cloud--cloud, cloud--hard and hard--hard contractions contribute the real exponent weights \(-f_\omega^2\), \(2f_\omega\) and \(-1\); the cloud phases cancel the virtual Coulomb phase \cite{sm-Choi:2017ylo}.

\section{Fixed soft emission configuration}
\label{SM:expansions}

For eqs.~\lettereq{PRL:stringratio} and \lettereq{PRL:radiationlog} of the Letter, the hard scattering lies in the \(x\)-\(z\) plane, while the physical soft graviton is emitted along \(+y\) with frequency \(\nu\), perpendicular to this plane. This differs from the forward emission in eq.~\lettereq{PRL:radiation-two-phases}. In the all incoming convention, the four point reference momenta and the chosen polarization are
\begin{equation}
 \begin{aligned}
 p_1&=E(1,0,0,1),&p_2&=E(1,0,0,-1),\\
 p_3&=-E(1,\sin\Theta,0,\cos\Theta),&
 p_4&=-E(1,-\sin\Theta,0,-\cos\Theta),\\
 k&=-\nu(1,0,1,0),&
 e&=\frac{(0,1,0,i)}{\sqrt2},\quad \epsilon_{\mu\nu}=e_\mu e_\nu .
 \end{aligned}
 \label{SM:kinematics}
\end{equation}
\noindent The minus sign in \(k\) denotes an outgoing graviton. We take small negative \(\Theta\) in the LE limit, so \(p_1+p_3\) has a positive transverse \(x\) component. Here \(s=4E^2\) and \(t=-4E^2\sin^2(\Theta/2)\). The tree soft theorem is \(\mathcal M_5=(S^{(0)}+S^{(1)})\mathcal M_4+O(\nu)\) \cite{sm-CachazoStrominger:2014,sm-Sen:2017soft}. Contracting the four external legs with this polarization, with the momentum derivatives evaluated before imposing the reference kinematics, gives
\begin{equation}
 \begin{aligned}
 S^{(0)}&=\frac{i\kappa E}{\nu}e^{-i\Theta}\sin\Theta
 \overset{\rm LE}{\simeq}-\frac{i\kappa q}{\nu},\\
 S^{(1)}&=\frac{\kappa e^{-i\Theta}}2
 (t\partial_t-s\cos\Theta\,\partial_s)
 \overset{\rm LE}{\simeq}\frac\kappa4
 (q\partial_q-2s\partial_s).
 \end{aligned}
 \label{SM:exactoperators}
\end{equation}
\noindent Acting on the Regge amplitude in eq.~\lettereq{PRL:string} of the Letter gives eq.~\lettereq{PRL:stringratio}.

At logarithmic accuracy, the operator \(S^{(1)}\) in eq.~\eqref{SM:exactoperators} acts on the full \(K_{\rm gr}^{\rm reg}\) from loop momenta above \(\nu\) \cite{sm-Sahoo:2018lxl}. With \(u=-s-t\), its analytic form and the physical boundary value at \(s+i0\) give
\begin{equation}
 \begin{aligned}
 K_{\rm gr}^{\rm reg}
 &=-\frac{2G_N}{\pi}\big[s\log(-s)+t\log(-t)+u\log(-u)\big]\log\nu^{-1},\\
 S^{(1)}K_{\rm gr}^{\rm reg}
 &=-\frac{\kappa G_N}{\pi}e^{-i\Theta}
 \left[t\log\frac{t}{u}-s\cos\Theta\big(\log(-s)-\log(-u)\big)\right]\log\nu^{-1}\\
 &\overset{\rm LE}{\simeq}-i\kappa G_Ns\,e^{-i\Theta}\cos\Theta\log\nu^{-1},\\
 -i\nu R_s\log\nu^{-1}\,S^{(0)}
 &=\kappa G_Ns\,e^{-i\Theta}\sin\Theta\log\nu^{-1}.
 \end{aligned}
 \label{SM:radiationlog}
\end{equation}
\noindent The imaginary part of \(K_{\rm gr}^{\rm reg}\) comes only from same-side pairs, while all six distinct pairs contribute to its real part. The LE limit suppresses the remaining terms in \(S^{(1)}K_{\rm gr}^{\rm reg}\).

The last line comes from three-graviton diagrams with loop momenta below \(\nu\): the emitted graviton exchanges a softer graviton with the outgoing hard particles. After absorbing the observer-distance phase into retarded time, their total energy \(2E\) gives \(2G_N(2E)\nu=\nu R_s\). The additional real term in this region vanishes for this transverse emission configuration. Since \(S^{(0)}\propto\nu^{-1}\), multiplication by the \(\nu\log\nu^{-1}\) rescattering correction gives a logarithmic subleading term. Thus \(S_{\rm gr}^{(1)}\) in the Letter combines both loop momentum regions, whereas \(S^{(1)}\) in eq.~\eqref{SM:exactoperators} is the tree differential operator,
\begin{equation}
 \begin{aligned}
 S_{\rm gr}^{(1)}
 &=S^{(1)}K_{\rm gr}^{\rm reg}-i\nu R_s\log\nu^{-1}\,S^{(0)}\\
 &\overset{\rm LE}{\simeq}-\frac{i\kappa^3s}{32\pi}e^{-i\Theta}
 (\cos\Theta+i\sin\Theta)\log\nu^{-1}.
 \end{aligned}
 \label{SM:radiationlog-result}
\end{equation}
\noindent This gives the classical \(\cos\) and \(i\sin\) terms of eq.~\lettereq{PRL:radiationlog}. Finally, restoring the undetermined hard reference scale replaces \(\log\nu^{-1}\) by \(\log(\nu_0/\nu)\); changing \(\nu_0\) changes only finite subleading terms.

\section{Gravitational radiation and rescattering}
\label{SM:radiation}

We derive eq.~\lettereq{PRL:radiation-two-phases} from the phases in eq.~\lettereq{PRL:rescattering}, using their transverse coordinates and \(2R_s\ll b\ll\nu^{-1}\), \(\theta\ll1\). The change in hard energy and the exchange between the emitted graviton and the other hard particle give the phase mismatch \cite{sm-Ciafaloni:2015unified}
\begin{equation}
 \begin{aligned}
 \Phi_R(\vec x)&=\log\frac{|\vec b-\vec x|}{b}
 -\frac E\nu\log\frac{|\vec b-(\nu/E)\vec x|}{b},\\
 \delta_R-\delta_{\rm in}&=-\nu R_s\Phi_R(\vec x).
 \end{aligned}
 \label{SM:phase-mismatch}
\end{equation}
\noindent The tree emission kernel in Fig.~\ref{PRL:fig:rescattering} contains the factor \(-R_s\Phi_R\), combining emission from the hard lines and the exchanged graviton. Summing its insertion over all ladder and crossed ladder orderings supplies
\begin{equation}
 \sum_{n=0}^{\infty}\frac{(2i)^n}{(n+1)!}
 \sum_{j=0}^{n}\delta_R^{\,j}\delta_{\rm in}^{\,n-j}
 =\frac{e^{2i\delta_R}-e^{2i\delta_{\rm in}}}
 {2i(\delta_R-\delta_{\rm in})}.
 \label{SM:insertion-sum}
\end{equation}
\noindent Multiplying by \(-R_s\Phi_R\) gives \((e^{2i\delta_R}-e^{2i\delta_{\rm in}})/(2i\nu)\). Removing the common elastic phase and expanding the incoming displacement to leading order in \(\nu/E\), while retaining the rescattering exponential, gives the transverse emission integral \cite{sm-Ciafaloni:2019infrared}
\begin{equation}
 2i\nu\mathfrak M_\lambda
 =\frac{\sqrt{G_Ns}}{2\pi^2}\int\frac{\dd^2\vec x}{r^2}
 e^{i\nu\vec\theta\cdot\vec x+i\lambda(\phi_\theta-\phi_x)}
 \left[\left(\frac{|\vec b-\vec x|}{b}\right)^{-2i\nu R_s}
 -e^{-i\nu\vec x\cdot\vec\Theta}\right],
 \label{SM:emission-integral}
\end{equation}
\noindent where \(r=|\vec x|\), \(\phi_v=\arg\vec v\), and \(\lambda=\pm2\). The prefactor and helicity phase already include the tree emission vertex. The angular integral is \(\int\dd\phi_x\,e^{i\lambda(\phi_\theta-\phi_x)+i\vec v\cdot\vec x}/(2\pi)=-e^{i\lambda(\phi_\theta-\phi_v)}J_2(|\vec v|r)\). Using \(|\vec b-\vec x|\simeq r\) in the logarithmically enhanced large-\(r\) region yields
\begin{equation}
 \frac{2\pi i\nu}{\sqrt{G_Ns}}\mathfrak M_\lambda
 \simeq\int_b^\infty\frac{\dd r}{r}
 \left[e^{i\lambda\psi}J_2(\nu r|\vec\theta-\vec\Theta|)
 -J_2(\nu\theta r)\left(\frac rb\right)^{-2i\nu R_s}\right].
 \label{SM:radial-emission}
\end{equation}
\noindent Since \(b\nu\theta,b\nu|\vec\theta-\vec\Theta|\ll1\), extending the lower endpoint to zero changes only terms beyond this accuracy. The two integrals are
\begin{equation}
 \begin{aligned}
 \int_0^\infty\frac{\dd r}{r}J_2(\nu r|\vec\theta-\vec\Theta|)&=\frac12,\\
 \int_0^\infty\frac{\dd r}{r}J_2(\nu\theta r)\left(\frac rb\right)^{-2i\nu R_s}
 &=\frac{\Gamma(1-i\nu R_s)}{2(1+i\nu R_s)\Gamma(1+i\nu R_s)}
 \left(\frac{b\nu\theta}{2}\right)^{2i\nu R_s}.
 \end{aligned}
 \label{SM:Bessel-Mellin}
\end{equation}
\noindent The ratio \(\Gamma(1-i\nu R_s)/[(1+i\nu R_s)\Gamma(1+i\nu R_s)]=1+O(\nu R_s)\) has no enhanced logarithm. Keeping \((b\nu\theta/2)^{2i\nu R_s}=e^{-2i\nu R_s\log[2/(b\nu\theta)]}\) and dropping the unenhanced terms gives eq.~\lettereq{PRL:radiation-two-phases}. The omitted angular corrections and finite terms remain outside this approximation. In the leading soft limit, the normalization used here is related to the standard soft factor by \(S^{(0)}_{\rm gr}\simeq2i(2\pi)^{3/2}\mathfrak M_\lambda\) for the same helicity convention.

\input{supplement_references}

\clearpage
\endgroup

\end{document}